\documentclass[preprint,secnumarabic,amssymb, amsmath,nobibnotes, aps, prb,showkeys]{revtex4}
\usepackage{epsfig}
\usepackage{latexsym}
\usepackage{graphicx}
\usepackage{indentfirst}
\begin{document}

\title{{Graphene nanoflakes - structural and electronic properties}}
\author{A.~Kuc and T.~Heine}
\address{School of Engineering and Science, Jacobs University Bremen, Campus Ring 1, 28759 Bremen, Germany}
\author{G.~Seifert}
\address{Physikalische Chemie, Technische Universit\"at, Bergstrasse 66b, 01062 Dresden, Germany}

\begin{abstract}

The structures, cohesive energies and HOMO-LUMO gaps of graphene nanoflakes and corresponding polycyclic aromatic hydrocarbons for a large variety of size and topology are investigated at the density functional based tight-binding level.
Polyacene-like and honeycomb-like graphene nanoflakes were chosen as the topological limit structures.
The influence of unsaturated edge atoms and dangling bonds on the stability is discussed.
Our survey shows a linear trend for the cohesive energy as function of N$_{\rm s}$/N (N $-$ total number of atoms and N$_{\rm s}$ is number of edge atoms).
For the HOMO-LUMO gap the trends are more complex and include also the topology of the edges.

\end{abstract}
\keywords{Graphene Nanoflakes, Polycyclic Aromatic Hydrocarbons, Topology Trends, Energetic Stability, Electronic Properties, DFTB}
\maketitle

\section{Introduction}

Graphene, as a single layer of carbon atoms arranged packed densely in a two-dimensional honeycomb crystal lattice, has attracted an enormous interest in the area of solid state electronics and composite materials, due to its high mechanical, thermal and chemical stability and excellent electronic properties.

Graphene nanoribbons (GNRs) and nanoflakes (GNFs), that are finite in both dimensions, can be considered as fragments or molecular subunits of graphene.
Since their initial successful fabrication,\cite{Novoselov2004} the dimensions of GNRs and GNFs have rapidly reduced from the microscale down to nanometer sizes either by top-down\cite{Li2008} or bottom-up\cite{Yang2008, Wang2008} approaches.
This gave a possibility to explore low-dimensional transport and perspective for carbon-based nanoelectronics.

Depending on the size and shape, GNFs possess the ability to form ordered columnar me\-so\-phases.\cite{Mueller2007}
Since the basic functional components of future electronics and spiroelectronics devices are required to be on the nanometer scale it is important to understand the properties of GNFs and their saturated counterparts, polycyclic aromatic hydrocarbons (PAHs).

Isolated GNFs and GNRs can be presently produced using different experimental approaches.
The bottom-up approach by thermal annealing nanographene molecules results in a conductive graphene film.\cite{Wang2008}
Alternatively, nanographene can be produced by soft-landing of ions generated by solvent-free matrix assisted laser desorption/ionisation.\cite{Rader2006}
The product is transfered to the gas phase, purified and adsorbed at surfaces.
As top-down-techniques, GNRs with widths varying from several tens of nanometers down to 2~nm have been fabricated either by etching\cite{Lemme2007, Campos2009, Ci2009} or by means of chemical treatment of graphene or graphite.\cite{Li2008}
It has been reported that GNRs with certain edge chirality would open the band gap.\cite{Li2008a}

GNFs with controlled thicknesses have been isolated in solution using density gradient ultracentrifugation.\cite{Green2009}
Cong et al\cite{Cong2009}.\ have fabricated arrays of graphene nanodiscs (GNDs) using nanosphere lithography (GNDs are GNFs with smooth edges and spherical shape).
Fabrication of GNRs with smooth edges is essential for many applications, however, it is difficult to produce such edges by conventional physico-chemical methods.
Jia et al.\cite{Jia2009}\ have shown that an efficient edge-reconstruction process, at the atomic scale, can be obtained for graphitic nanoribbons by Joule heating using an integrated transmission electron microscope-scanning tunnelling microscope (TEM-STM) system.

Very recently, few groups have independently reported very elegant methods for GNR production.\cite{Tapaszto2008, Hirsch2009, Jiao2009}
The methods are based on the longitudinal unzipping of multi-walled CNTs and involve an Ar plasma etching of the nanotubes.
These procedures are simple and inexpensive, and lead to GNRs with well-defined widths and edge structures.

It is widely known that stable GNFs with sizes C$_3$--C$_{10}$ form linear conformations, C$_{11}$--C$_{20}$ form annular structures and C$_{\rm n}$s with ${\rm n}>20$ form fullerenes.\cite{Jones1997, Zerbetto1999, Heine2002}
However, it appears that so far little is known on the stability, structure and properties of larger (planar) carbon nanostructures, and a computational study is timely.

Graphite and graphene are zero-gap semiconductors.
If carbon particles are reduced in size to a level where quantum effects are significant, large energy gaps may appear.\cite{Hunt1998}
Indeed, as large as 8.5 eV energy gaps have been predicted for clusters with few carbon atoms,\cite{Liang1990} while gaps below 2 eV were found for carbon cages composed of up to 80 carbon atoms.\cite{Saito1991,Woo1993,Zhang1992}
The knowledge and understanding of the size-tuned properties, e.g.\ binding energy and HOMO-LUMO gap, would make carbon particles interesting candidates for applications in nanotechnology.
 
\begin{figure}[ht!]
\begin{center}
\includegraphics[scale=0.50,clip]{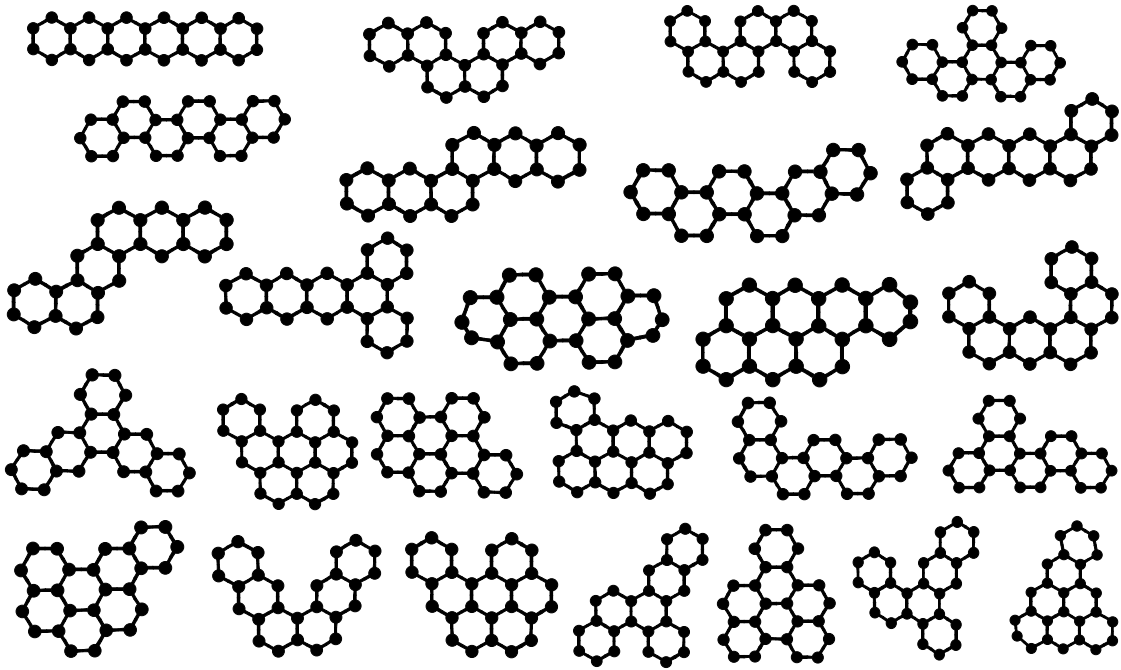}
\vspace*{-0.5cm}
\caption{\label{fig:1a}\footnotesize{Exemplary graphene nanoflakes isomers considered for C$_{\rm 26}$.}}
\end{center}
\end{figure}
In this paper, we focus on small and medium neutral GNFs and corresponding PAHs.
We have restricted our calculations to even number of atoms, as well as to planar sixfold ring systems (polyhexe structures; see Fig.~\ref{fig:1a}).
The determination of topologically distinct structures (isomers) is difficult, as the number of isomers is strictly related to the number of atoms ($n$) and increases rapidly with increasing $n$.
As an example, for the molecular formula C$_6$H$_6$, about 330 isomers can be written, considering the geometrical- and stereo-isomers.\cite{Dinadayalane2004}
The complete set of isomers for any graphene flage can be created using graph theory.\cite{Randic2003, Estrada2004}
In this work, we do not aim to investigate all possible graphene flake structures, instead we want to derive trends in stability and electronic structure on the basis of an extensive survey of flake structures inhibiting different molecular weight and topology.
We have created a number of possible isomers for structures with $N$=6--34, 38, 42, 50, 54, 60, and 74 ($N$ is the number of carbon atoms).
For some specific topologies, e.g.\ triangles, nanoribbons (stripes) or circular flakes, the number of atoms was up to 220.
Since the number of isomers rapidly increases with increasing $N$ the completeness in selection was related only to the smallest structures ($N<$28).

\section{Methods}
\label{Sec:Methods}

\begin{figure}[ht!]
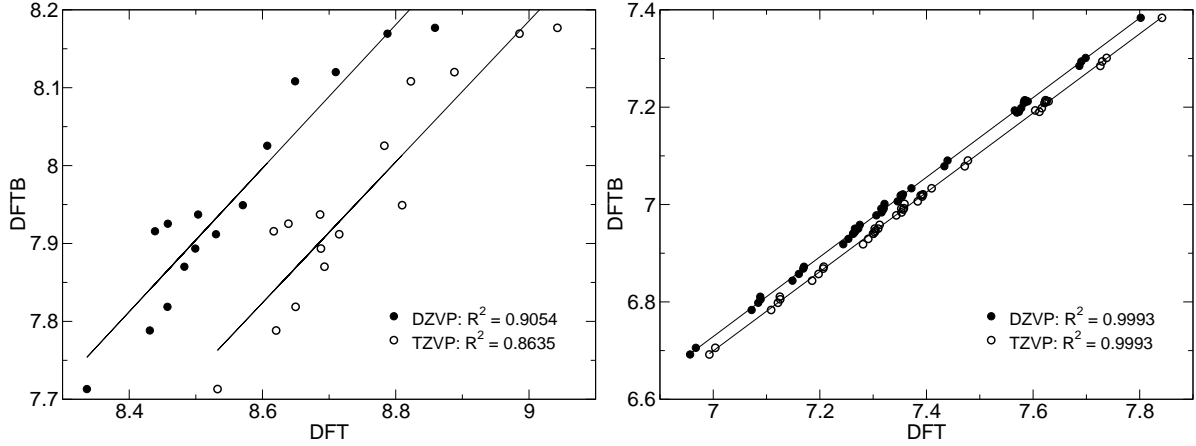

\begin{center}
\includegraphics[scale=0.30,clip]{2a.eps}
\includegraphics[scale=0.30,clip]{2b.eps}
\vspace*{-0.5cm}
\caption{\label{fig:2}{\footnotesize Comparison of the DFTB and the DFT cohesive energies for carbon flakes: (left) GNFs and (right) PAHs. The energies are given in units of the graphene cohesive energy.}}
\end{center}
\end{figure}

All structures and corresponding cohesive energies were calculated using DFTB (Density Functional based Tight-Binding).\cite{Porezag1995,Seifert1996}
For a recent review on the method see Oliveira et al.\cite{Oliveira2009}
To proof the reliability of the this method we have performed DFT (Density Functional Theory) calculations as well.
The VWN~\cite{Vosko1980} (Vosko, Wilk, Nusair) exchange-correlation potential was used for full DFT optimization with DZVP and TZVP basis sets.\cite{Godbout1992}
Both DFTB and DFT results are compared in Fig.~\ref{fig:2}, and we find good agreement between the two methods.
 
\begin{figure}[ht!]
\begin{center}
\includegraphics[scale=0.30,clip]{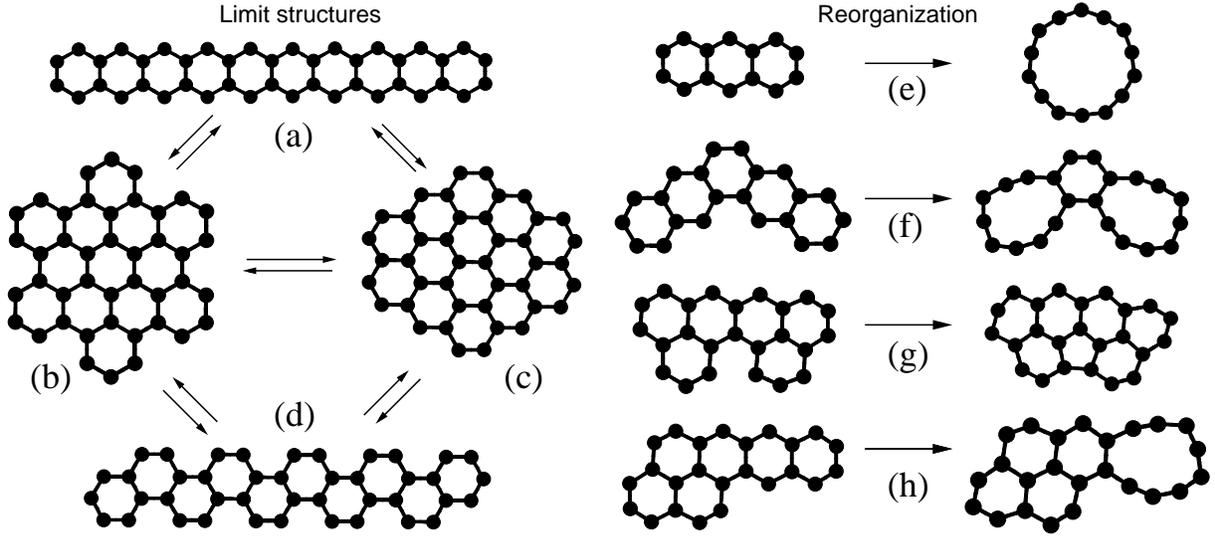}
\vspace*{-0.5cm}
\caption{\label{fig:1}\footnotesize{Considered limiting geometries for C$_{42}$ GNFs: (a) -- polyphenanthrenes (armchair nanoribbons) and (d) -- polyacenes (zig-zag nanoribbons), (b) -- armchair and (c) -- zig-zag circular flakes. Exemplary GNFs with unstable polyhexe forms before and after full geometry optimization: (e) C$_{14}$, (f) C$_{18}$, (g, h) C$_{22}$.}}
\end{center}
\end{figure}

For a given number of carbon atoms the limiting geometries were chosen: polyacene and polyphenanthrene chains,  zig-zag and armchair honey-comb-like flakes.
The exemplary structures for C$_{42}$ are shown in Fig.~\ref{fig:1}.

In order to examine the the size-dependent trends of properties for various GNF and PAH isomers, we have calculated the cohesive energies.
The cohesive energies were further related to the energy of the reference system -- a graphene layer.
One may write the binding energy $E_{\rm{bind}}$ (per atom) of nanoflakes  as:
\begin{equation}
\label{eq:1}
\frac{E_{\rm{bind(N)}}} {N} = \varepsilon_\infty + \gamma(N) 
\end{equation}
where $\varepsilon_\infty$ is the binding energy per atom in the infinite graphene layer and $\gamma(N)$ is a kind of a  surface or 'edge' energy.

For planar nanoflakes this surface energy can be expressed as a ratio of the number of edge atoms ($N_s$) to the total number of atoms ($N$):
\begin{equation}
\label{eq:2}
\frac{E(N)} {N} = \varepsilon_\infty + c\frac{N_s} {N},
\end{equation}
where $c$ is a constant.
In case of GNFs, the $N_s$ stands for the number of unsaturated carbon atoms, the atoms that have only two neighbors, while for PAHs it is directly related to the number of hydrogen atoms.

Since the number of atoms $N$ in a planar flake is proportional to $R^2$ ($R$ can be a radius for circular flakes or one half of the diagonal for other shapes), then $\frac{N_s} {N}$ scales as $\frac{1} {\sqrt{N}}$ and energy of graphene nanoflakes can be written as:
\begin{equation}
\label{eq:3}
\frac{E(N)} {N} = \varepsilon_\infty + \frac{c'} {\sqrt{N}},
\end{equation}
where $c'$ is a constant.
Thus, the correlation between the binding energy and the size of nanoflakes should be proportional to $\frac{1} {\sqrt{N}}$ as well as proportional to $\frac{N_s} {N}$, if quantum effects of the extended $\pi$-system are not important.

\section{Results and Discussion}

\subsection{Structural Properties}

In this work, we have studied structural, energetic and electronic properties graphene nanoflakes (GNFs) and their saturated counterparts -- polycyclic aromatic hydrocarbons (PAHs).
The family of GNFs has dangling bonds at the circumference of the flakes which are saturated with hydrogen atoms in the case of PAHs.
Therefore, some of the GNF structures suffered reorganization of the atoms arrangement during the optimization, formation of  monocycles (MC), 'holes' or 5-fold rings (see Fig.~\ref{fig:1}(e-h)).
This effect is not observed in the PAH systems, what suggests the possibility to stabilize small graphene flakes in their polyhexe forms by hydrogenation.

Several specific topologies can be mentioned, among them circular flakes, triangles, nanoribbons.
The latter ones are characterized by the number of hexagonal units that determine the flake's width.
The nanoribbons, with the width of one hexagonal unit, are polyacenes and polyphenanthrenes, depending on the type of edges (zig-zag or armchair).
All GNFs in the polyacene form are unstable in their polyhexe forms and they transform into monocyclic rings or partial rings ('holes') after full optimization.
Therefore, we have included also monocyclic rings in our studies for comparison and validation of our method, as  these systems have been investigated intensively - see e.g. \cite{Bylaska1998, Saito1999, Torelli2000, Xu2006}

\begin{figure}[ht!]
\begin{center}
\includegraphics[scale=0.85,clip]{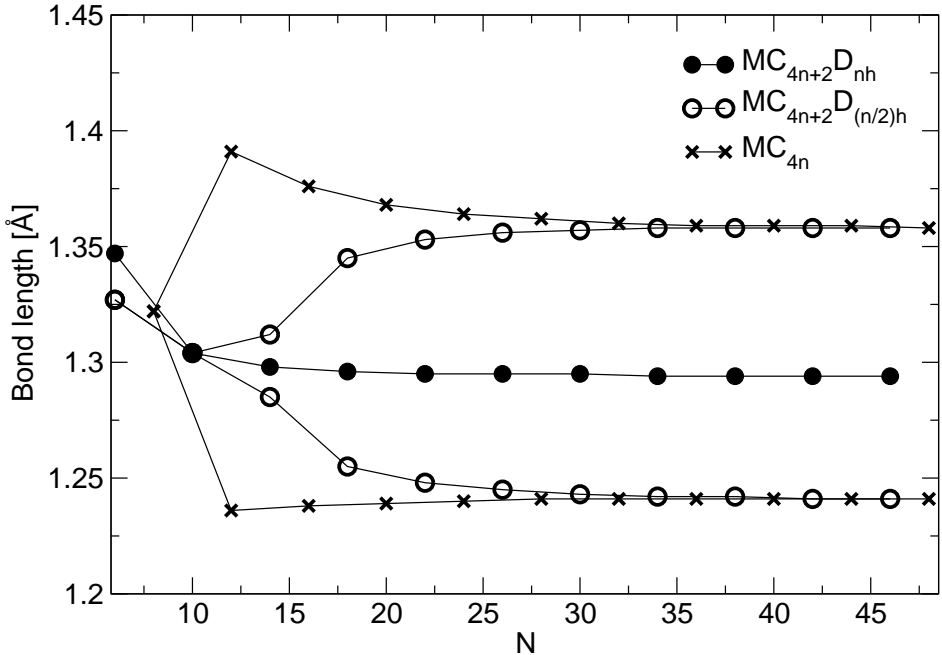}
\vspace*{-0.5cm}
\caption{\label{fig:4a}\footnotesize{Calculated bond lengths in the 4$n$+2 (with symmetry $D_{{\rm N}h}$ and $D_{{\rm (N/2)}h}$) and 4$n$ monocycles.}}
\end{center}
\end{figure}

We have divided carbon MCs into two families: 4$n$+2 (with symmetry $D_{{\rm N}h}$ and $D_{{\rm (N/2)}h}$, $N$ -- number of carbon atoms) and 4$n$ ($n$ -- natural number), following the H\"uckel rule of aromaticity.
The 4$n$+2 MCs with symmetry $D_{{\rm N}h}$ are called cumulenic and have all bond lengths and bond angles equal.
MCs with $D_{{\rm (N/2)}h}$ symmetry are those with alternating bond lengths (or bond angles).
The bond lengths of all studied carbon MCs are shown in Fig.~\ref{fig:4a}.

Our results and the previous theoretical studies show that the energetically most stable 4$n$ MCs have always bond-lengths alternant structures.\cite{Bylaska1998, Saito1999, Torelli2000, Xu2006}
This results from the first-order Jahn-Teller distortion but the alternation decreases with increasing the ring size.
Furthermore, we have found that the 4$n$+2 ground state structures prevent the bond-length alternation up to $N$ = 10, while larger rings show a bond length alternation ($D_{{\rm (N/2)}h}$) at all sizes and the cumulenic isomer ($D_{{\rm N}h}$) is a structural transition state.
This results are in good agreement with other works at DFT and quantum Monte Carlo level of calculations,\cite{Bylaska1998, Torelli2000} however, the energy difference between cumulenic and alternant structure of C$_{\rm 10}$ is extremely small of $\simeq$1~kcal~mol$^{\rm -1}$.

\subsection{Energetic Stability}

\begin{figure}[ht!]
\begin{center}
\includegraphics[scale=0.65,clip]{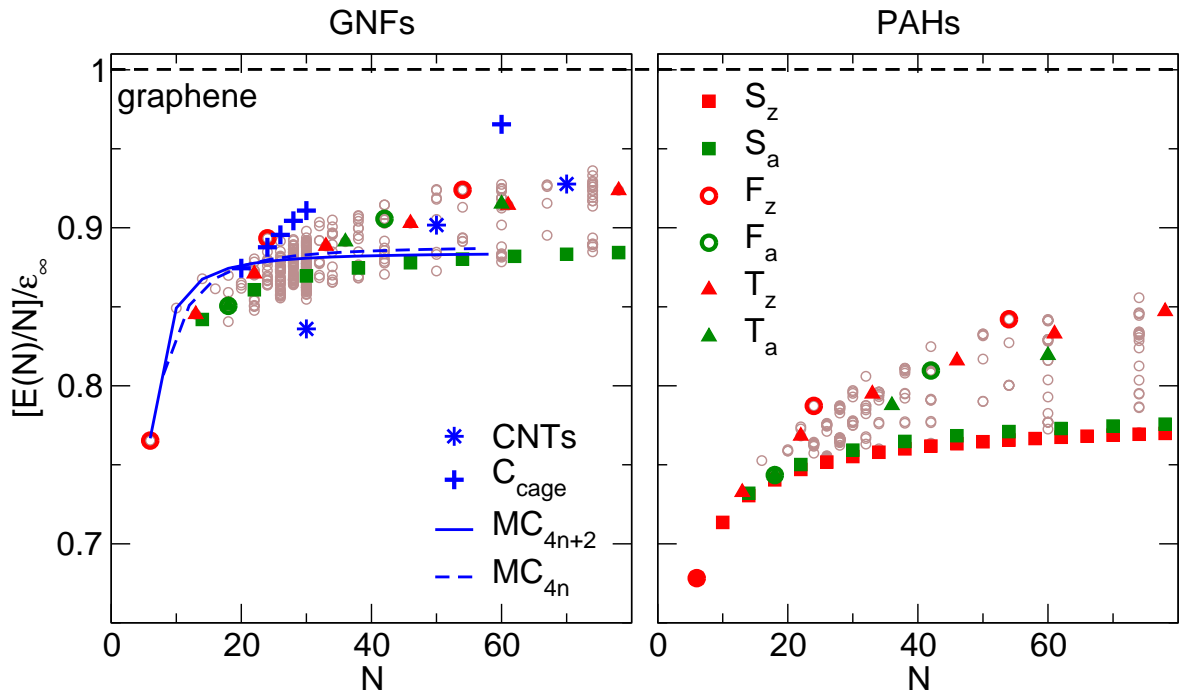}
\vspace*{-0.5cm}
\caption{\label{fig:4}\footnotesize{(Online color) Calculated cohesive energies (in units of the graphene cohesive energy, $\varepsilon_\infty$) of GNFs and PAHs versus the number of carbon atoms ($N$). Some of the topologies together with the type of edges (z -- zig-zag or a -- armchair) are highlighted: (squares) S -- stripes (nanoribbons), (circles) F -- circular flakes, (triangles) T -- triangles, MC -- monocycles. For comparison finite carbon nanotubes (CNTs; stars) and fullerenes (carbon cages; pluses) are given.}}
\end{center}
\end{figure}

The calculated cohesive energies of all studied GNF and PAH isomers as a function of the number of carbon atoms are given in Fig.~\ref{fig:4}.
In addition, cohesive energies of other carbon allotropes (fullerenes and carbon nanotubes, CNTs) are shown.
Some of the high-symmetry topologies (triangles, circles, nanoribbons) are marked as well.

The most stable topologies,for a given $N$, are circular flakes, while the least stable are the very narrow nanoribbons, namely polyacenes and polyphenanthrenes.
In case of GNFs, the polyacene form transforms into monocycles or 'holes', therefore the least stable structures will be those with partial polyacene forms.
Nanoribbons become more stable with increasing their width approaching the stability of graphene for very wide systems.
In fact, the most and least stable topologies correspond to our suggested limiting structures.
Other energetically favorable topologies are those of triangular flakes.
Recently, it was shown by Ci et al\cite{Ci2009}.\ that graphene fragments can be shape-controlled by multistage cutting and that the two main shapes obtained in such a process are triangles and few-nanometer wide nanoribbons.

\begin{figure}[ht!]
\begin{center}
\includegraphics[scale=0.45,clip]{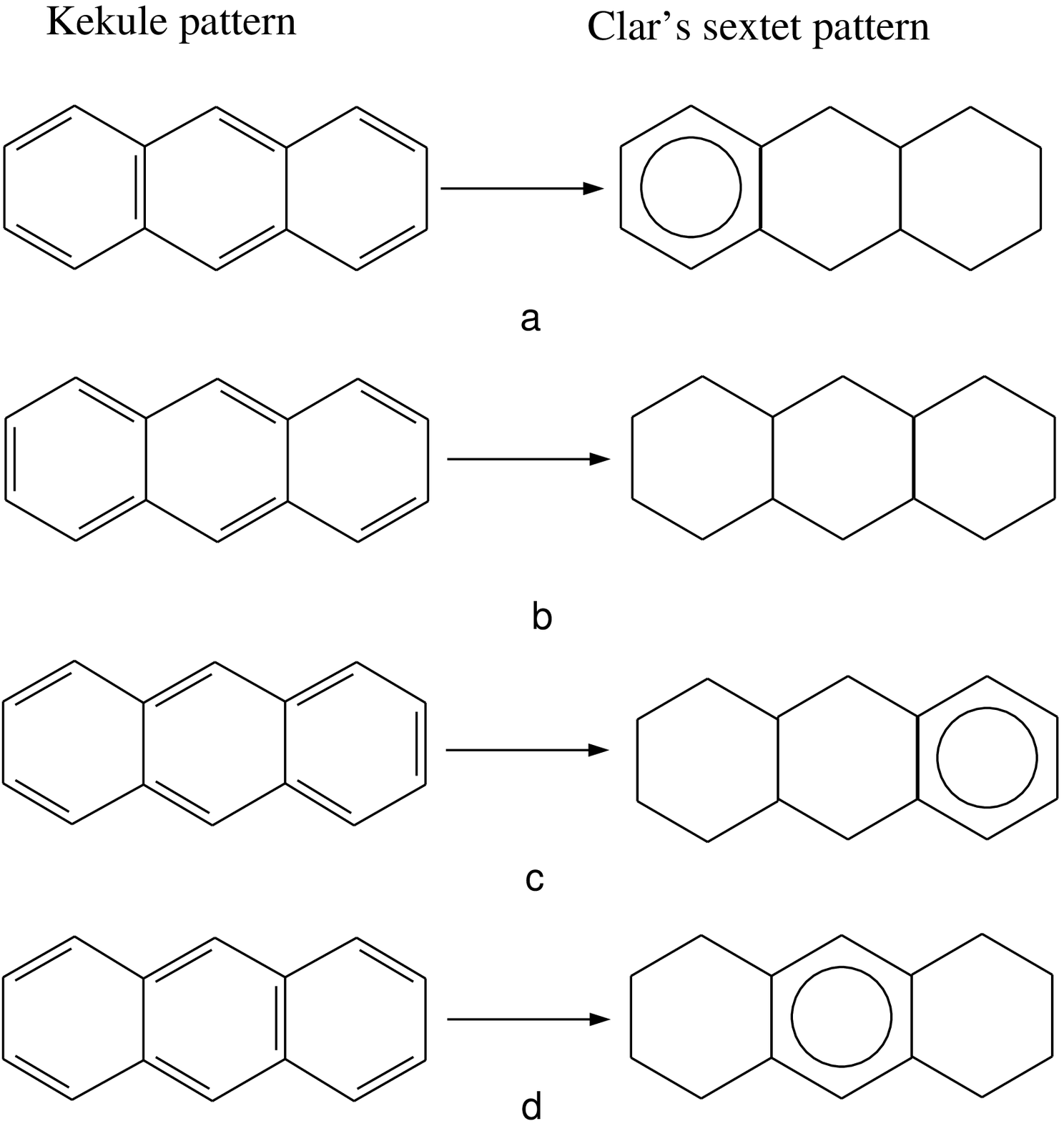}
\includegraphics[scale=0.65,clip]{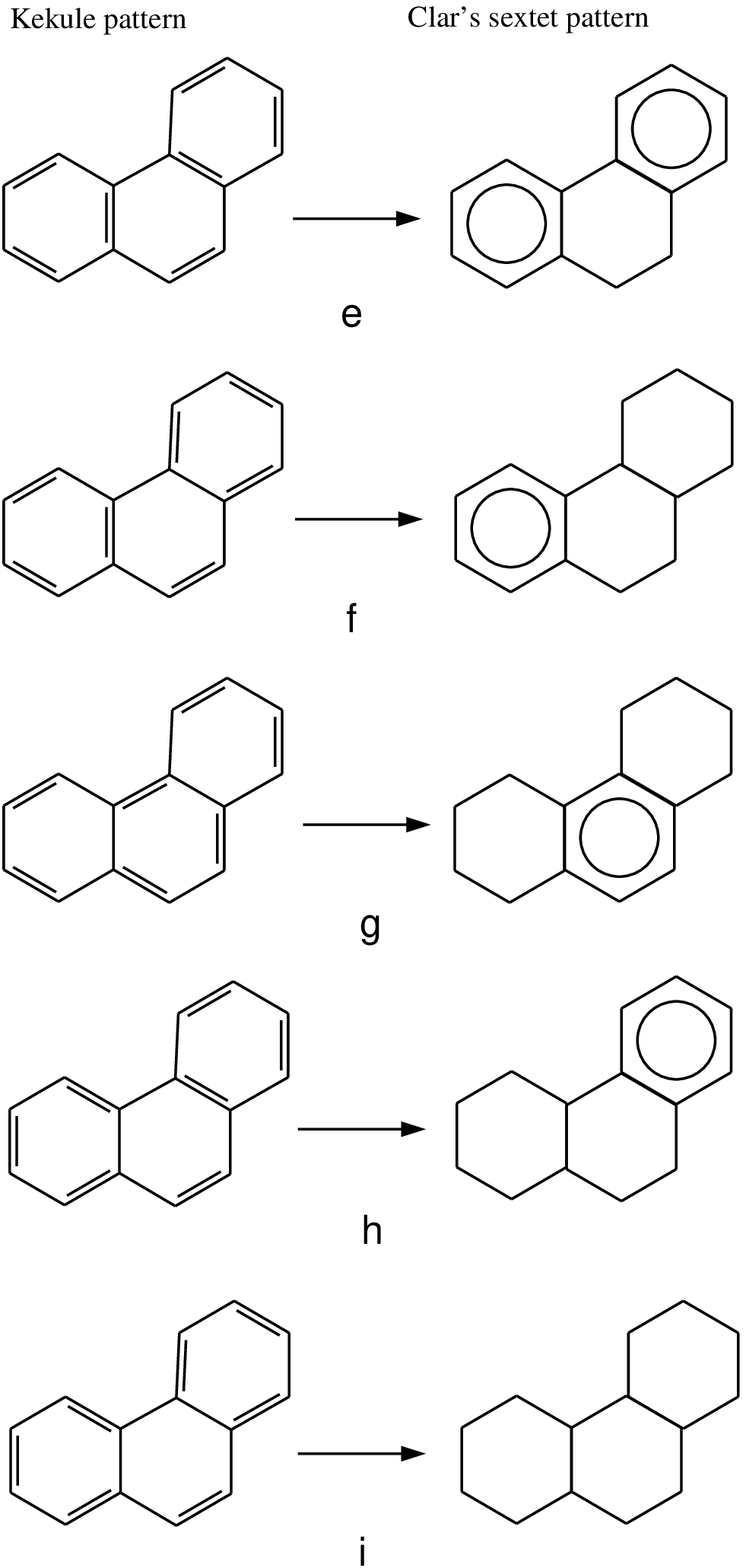}
\vspace*{-0.5cm}
\caption{\label{fig:KP}\footnotesize{Kekul\'{e} and corresponding Clar's sextet patterns of anthracene (a-d) and phenanthrene (e-i).}}
\end{center}
\end{figure}

Generally, the zig-zag type of edges is more stable than the armchair one.
However, in the case of nanoribbons, polyphenanthrenes are more stable than polyacenes, what can be seen in the case of PAHs.
This result is in qualitative agreement with other theoretical studies.\cite{Wu2000}
The reason for the difference in the stabilities can be attributed to the differences in the geometric properties, namely to the number of the Kekul\'{e} numbers, which is larger for armchair edges.
Considering C$_{\rm {14}}$H$_{\rm {10}}$ one can draw 5 Kekul\'{e} patterns for phenanthrene but only 4 patterns for anthracene (see Fig~\ref{fig:KP}).

The set of Kekul\'{e} patterns can be corresponded to the set of aromatic sextet patterns according to Clar's notation of the $\pi$-sextets.\cite{Clar1972, Ohkami1981}
According to Clar, not all of the Kekul\'{e} valence structures in PAHs are equally important, and benzene may be the only system that is exceptional.
The dominant ones are those Kekul\'{e} valence structures which, after superimposing, will give the largest number of isolated $\pi$-sextet rings.\cite{Randic1998}
Therefore, PAHs larger than benzene tend to form with maximum number of aromatic sextets.
Fig~\ref{fig:KP} shows that for polyacenes we can draw at maximum one Clar sextet, while for polyphenanthrenes this number increases with the length.

All the GNFs are less stable than graphene layer and the isolated C$_{\rm {60}}$ molecule, but are as stable as smaller fullerenes and (5,5) CNTs with finite length.
In fact, it was previously shown that for N$<$30 the graphene-like clusters are more stable than the respective fullerene structures.\cite{Jones1997, Jones1999,Seifert1998}

Considering carbon MCs, we have found that they are more stable than carbon flakes up to $N$ = 20 and this is in agreement with other theoretical works.\cite{Jones1997, Jones1999}
Polyacene GNFs are not stable in their polyhexe forms for smaller sizes and they reorganize into MCs, what is connected with the overlapping of free p-orbitals at the edges of GNF planes due to dangling bonds.
This effect causes a gain in the binding energy.

\begin{figure}[ht!]
\begin{center}
\includegraphics[scale=0.60,clip]{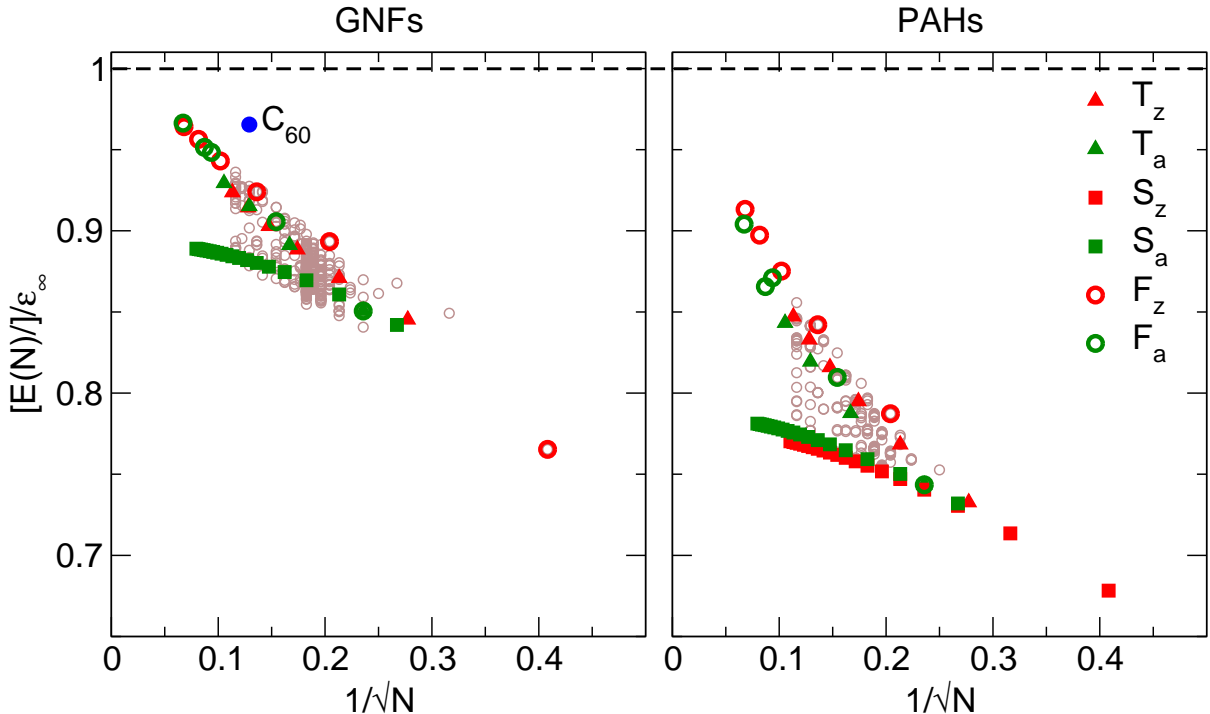}
\vspace*{-0.5cm}
\caption{\label{fig:6}\footnotesize{(Online color) Calculated cohesive energies (in units of the graphene cohesive energy, $\varepsilon_\infty$) of GNFs (left) and PAHs (right) versus $\frac{1}{\sqrt{N}}$ ($N$ -- total number of carbon atoms). For the abbreviations see Fig.~\ref{fig:4}.}}
\end{center}
\end{figure}
According to Eq.~\ref{eq:3} (see Sec.~\ref{Sec:Methods}), the cohesive energy should increase with $\frac{1}{ \sqrt{N}}$ if the quantum effects of the $\pi$ system have no major influence on the stability.
Figure~\ref{fig:6} shows the cohesive energy plot as a function of $\frac{1}{ \sqrt{N}}$.
For a given topology, e.g.\ triangles, stripes, circular flakes, and flake sizes $N>$18 there are linear trends of cohesive energies.
However, the overall behavior is far from linearity.
The energies of circular and triangular flakes converge to the energy of graphene layer, while the narrow nanoribbons converge to the energies of infinite polyacene and polyphenanthrene.
The same trend is observed for both, GNFs and PAHs (see Fig.~\ref{fig:6}).

\begin{figure}[t]
\begin{center}
\includegraphics[scale=0.55,clip]{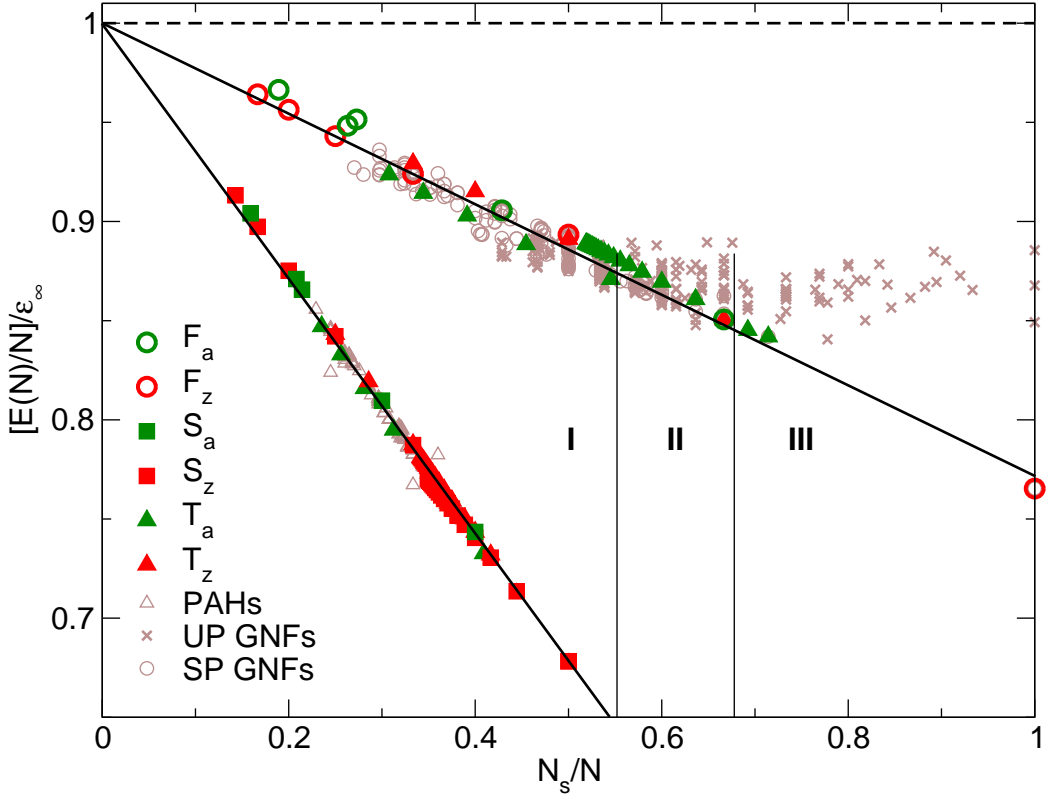}
\vspace*{-0.5cm}
\caption{\label{fig:8}\footnotesize{(Online color) Calculated cohesive energies (in units of the graphene monolayer cohesive energy, $\varepsilon_\infty$) of GNFs and PAHs versus $\frac{N_s}{N}$ ($N_s$ -- number of edge atoms, $N$ -- total number of carbon atoms). For the abbreviations see Fig.~\ref{fig:4}.}}
\end{center}
\end{figure}

As one can see from Fig.~\ref{fig:8} there is a much better linear correlation of the binding energies from the $\frac{N_s}{N}$ ratio for GNFs and PAHs.

In the case of PAHs, the benzene molecule can be taken as the bottom limiting structure with the ratio $\frac{N_H} {N_C+N_H}$ equal to 0.5.
The top limit stands, of course, for the infinite graphite monolayer with the ratio $\frac{N_H} {N_C+N_H}$=0.
One should keep in mind that this ratio should vary for other isomers between 0 and 0.5.

In Fig.~\ref{fig:8} the studied GNFs can be divided into two groups: (a) those, that keep the polyhexe form after geometry optimization and (b)those, where the polyhexe form of a flake is not stable.
Furthermore, one can distinguish also between three size regions in the cohesive energy plot.
In the $region~I$ only GNFs stable in the hexagonal forms are cumulated, while the systems that suffer strong reorganization in $\pi$-electron system are present in the $region~III$.
The intermediate structures are marked in the $region~II$.
The higher cohesive energies in the $region~III$ are due to the formation of monocyclic and 5-membered rings, where stronger overlap of p-orbitals appears.

Taking into consideration only stable forms of GNFs, the energy scales linearly with the size of a flake for all the topologies studied.
All the energies converge to that of graphene single layer, unlike for the $\frac{1}{ \sqrt{N}}$ ratio.
It can be concluded that the number of the edge atoms is very important for the measure of surface energy, $\gamma$($N$).
This can be understood better by the fact that the deviations in the energy of carbon flakes from that of graphene layer are mainly due to the edge atoms.
The points in the $region~I$ can be described by the trend line, whose slope gives $\gamma(N)$.

As one can also see from Fig.~\ref{fig:8}, a perfect linear trend of the cohesive energy versus $N_s/N$ is obtained for PAHs.
In this case, the surface energy, $\gamma(N)$, is a sum of the surface energy of carbon atoms ($\varepsilon_s$) and the surface energy of hydrogen atoms ($\varepsilon_H$).
While, for the GNFs (homonuclear molecules) it can be calculated in a clear way, there is no unique procedure to determine the binding energy in heteronuclear molecules.
Still one can make an approximation that $\varepsilon_s$ is the correction of $\varepsilon_\infty$ and should be equal zero in the case of PAHs, and then the $\varepsilon_H$ can be calculated.
In conclusion, small graphene flakes can be stabilized in their polyhexe forms by saturation of dangling bonds by hydrogen atoms.

\subsection{Electronic Properties}

We have also studied the electronic properties of carbon flakes in terms of their size and topology.
The HOMO-LUMO gaps ($\Delta$) were calculated for all flakes studied in this work.
In addition, we have calculated much larger flakes (up to $N$ = 220) to compare the results with the experimental work of M\" uller et al.\cite{Mueller2007}

\begin{figure}[ht!]
\begin{center}
\includegraphics[scale=0.55,clip]{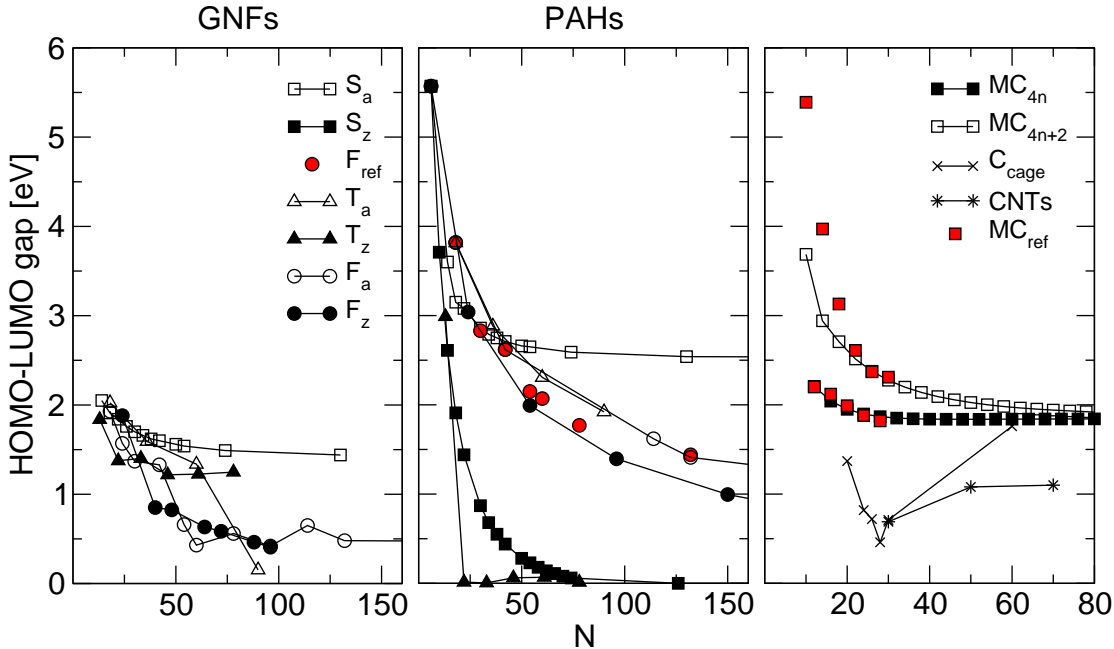}
\vspace*{-0.5cm}
\caption{\label{fig:9}\footnotesize{(Online color) Calculated HOMO-LUMO gap versus $N$. S -- stripes (polyacenes and polyphenanthrenes), F -- circular flakes, F$_{\rm ref}$ -- flakes as studied in Ref.~\onlinecite{Mueller2007}, T -- triangles, MC -- monocycles, MC$_{\rm ref}$ -- data from Ref.,~\onlinecite{Xu2006} z -- zig-zag, a -- armchair.}}
\end{center}
\end{figure}

Increasing the size leads to decrease of the HOMO-LUMO gap.
All nanoflakes studied in this survey are semiconductors or insulators (see Fig.~\ref{fig:9}).
However, large clusters tend to a gap closing, similar to graphene.
The size and shape dependent trends are divided into several groups of nanoflakes: polyacenes, polyphenanthrenes, monocycles, and graphene flakes of armchair and zig-zag edges.
Generally, for each specific topology, there is a size-dependent trend and the $\Delta$s decrease slowly or rapidly with the number of aromatic sextets, depending on the type of edges and the shape of flakes.

The $\Delta$ of polyphenantherene GNFs are lower than the corresponding PAH forms and they converge to the values of ~1.4 eV and ~2.5 eV, respectively.
When increasing the width of armchair-edged nanoribbons, however, the band gap approaches zero.\cite{Yoshizawa1998, Sharma2009}
Generally, the zig-zag type of peripheries lower the HOMO-LUMO gap in comparison to the armchair edges, as predicted by Stein et al.\cite{Stein1987}
Therefore, the gaps of polyphenanthrenes are larger than those of polyacenes.
The $\Delta$ of the PAH polyacenes decreases very rapidly with increasing the length and the width of nanoribbons, although a very small gap always occur.
This is due to a higher-order Peierls distortion effect.\cite{Kertesz1983}
In the zig-zag nanoribbons, the frontier orbitals are localized at the edges, while in the case of armchair-edge structures they are distributed evenly over the carbon structure.

We have found that the metallic character, for the range of sizes studied here, is found for the zig-zag-edge PAH structures with triangular topology.
The $\Delta$ decreases rapidly to zero already for $N$=20.
This is in a very good agreement with the recent work of Ezawa et al\cite{Ezawa2007}.\ who have found that the band gap decreases inversely to the length, and zero-energy states emerge as the length goes to infinity.
Infinite-length nanoribbons have the flat band made of degenerated zero-energy states.\cite{Ezawa2007}
The circular PAH flakes converge $\Delta$ to zero very slowly.
For example, the armchair flake with 222 carbon atoms has still $\Delta$ of around 1.1 eV.
The corresponding GNF structures do not show smooth trends and both types of edges, zig-zag and armchair, give similar values of $\Delta$.

Moreover, all carbon monocycles are semiconducting and the calculated $\Delta$ converge quickly to around 1.8 eV for both, 4$n$+2 and 4$n$, groups of MCs.
The variation of $\Delta$ is, however, much smaller for the 4$n$ MCs and faster convergence is obtained.
This result is in qualitative (for the 4$n$ MCs also quantitative) agreement with the work of Xu et al.\cite{Xu2006}

\section{Conclusions}

In this work, we have presented results of an extensive DFTB study on the structural and energetic properties of graphene nanoflakes and the corresponding polycyclic aromatic hydrocarbons.
A wide range of isomers (topologies) for a given number of carbon atoms, $N$, was considered ($N$ =6-34 (even numbers), 38, 42, 50, 54, 60, 67 and 74).
The results show clear trends in the energetic stabilities of different topological types with changing the flake size. 
We have focused on the energy trends of neutral flakes considering the total number of carbon atoms, as well as, the number of edge atoms.
Only six-fold ring connections in the planar clusters were considered in this study.

The established model consideration describes the cohesive energy trends in terms of the number of atoms qualitatively well.
For both, GNFs and PAHs, the energy scales as $\frac{1} {\sqrt{N}}$, however, different topologies converge to the energies of the corresponding infinite structures.
Only the triangular and circular flakes approach the energy of graphene reference structure.

Our simple model consideration predicts a linear behavior of cohesive energy versus the ratio $\frac{N_s}{N}$ (where $N$ is the total number of carbon atoms and $N_s$ denotes the number of edge atoms).
Good linear scaling is obtained for the all the PAHs and these GNFs, which are stable in their six-fold connections.
Small GNFs undergo a reorganization of $\pi$-electron system forming monocycles and 'holes', what causes a gain in energy and the deviations from the linearity.
At this point we must, however, emphasize that Eq.~\ref{eq:3} has been tested for few topologies.
It is not obvious if this model can be generalized to hold for the rich manifold of other graphene nanoflake topologies.

We have also found interesting size- and topology-dependent trends in the electronic properties.
Calculations of HOMO-LUMO gaps shows a variety of properties in electronic conduction, from metals to typical semiconductors and insulators.
Generally, the zig-zag type of edges lower the energy gap and almost all zig-zag triangular PAHs are metallic.
Armchair triangular and all circular PAHs approach the electronic properties of graphene very slowly.
GNFs do not show smooth trends of $\Delta$ but the $\Delta$ decreases with increasing number of stable aromatic sextets.

Comparison between the DFTB and DFT calculation shows a good agreement, validating our method for its usage for systems built of $sp^2$ carbon atoms.

\end{document}